\documentclass[aps,showpacs,pra,twocolumn,footinbib,superscriptaddress]{revtex4-1}
\usepackage{epsfig}
\usepackage{amsfonts}
\usepackage{amsmath}
\usepackage{bm}
\usepackage{xcolor}
\usepackage{booktabs}
\usepackage{newtxtext,newtxmath}
\usepackage{yfonts}

\usepackage[utf8]{inputenc}
\usepackage[colorlinks,bookmarks=false,citecolor=blue,linkcolor=blue,urlcolor=blue]{hyperref}
\usepackage{soul}
\usepackage{stackengine} 

\usepackage{changepage}

\newcommand{\bra}[1]{\left< #1 \right|}
\newcommand{\ket}[1]{\left| #1 \right>}

\newcommand{\av}[1]{\left< #1 \right>}

\newcommand{\primesum}{\sideset{}{'}\sum}

\newcommand{\be}{\begin{equation}}
\newcommand{\ee}{\end{equation}}

\begin{document}

\title{Long-range multi-body interactions and three-body anti-blockade in a trapped Rydberg ion chain}
\author{Filippo M. Gambetta}
\affiliation{School of Physics and Astronomy, University of Nottingham, Nottingham, NG7 2RD, United Kingdom}
\affiliation{Centre for the Mathematics and Theoretical Physics of Quantum Non-equilibrium Systems, University of Nottingham, Nottingham NG7 2RD,  United Kingdom}
\author{Chi Zhang}
\affiliation{Department of Physics, Stockholm University, 10691 Stockholm, Sweden}
\author{Markus Hennrich}
\affiliation{Department of Physics, Stockholm University, 10691 Stockholm, Sweden}
\author{Igor Lesanovsky}
\affiliation{School of Physics and Astronomy, University of Nottingham, Nottingham, NG7 2RD, United Kingdom}
\affiliation{Centre for the Mathematics and Theoretical Physics of Quantum Non-equilibrium Systems, University of Nottingham, Nottingham NG7 2RD,  United Kingdom}
\affiliation{Institut für Theoretische Physik, University of Tübingen, 72076 Tübingen, Germany}
\author{Weibin Li}
\affiliation{School of Physics and Astronomy, University of Nottingham, Nottingham, NG7 2RD, United Kingdom}
\affiliation{Centre for the Mathematics and Theoretical Physics of Quantum Non-equilibrium Systems, University of Nottingham, Nottingham NG7 2RD,  United Kingdom}

\date{\today}

\begin{abstract}
Trapped Rydberg ions represent a flexible platform for quantum simulation and information processing which combines a high degree of control over electronic and vibrational degrees of freedom. The possibility to individually excite ions to high-lying Rydberg levels provides a system where strong interactions between pairs of excited ions can be engineered and tuned via external laser fields. We show that the coupling between Rydberg pair interactions and collective motional modes gives rise to effective long-range and multi-body interactions, consisting of two, three, and four-body terms. Their shape, strength, and range can be controlled via the ion trap parameters and strongly depends on both the equilibrium configuration and vibrational modes of the ion crystal. By focusing on an experimentally feasible quasi one-dimensional setup of $ {}^{88}\mathrm{Sr}^+ $ Rydberg ions, we demonstrate that multi-body interactions are enhanced by the emergence of soft modes associated, e.g., with a structural phase transition. This has a striking impact on many-body electronic states and results, for example, in a three-body anti-blockade effect which can be employed as a sensitive probe to detect structural phase transitions in Rydberg ion chains. Our study unveils the possibilities offered by trapped Rydberg ions for studying exotic phases of matter and quantum dynamics driven by enhanced multi-body interactions.
\end{abstract}

\maketitle

\textit{Introduction.---} 
The coupling between internal atomic states and collective vibrational modes is the hallmark of trapped ion setups. The possibility to engineer phonon-mediated two-body interactions, which can be tuned via laser fields and trapping parameters, combined with single-ion control and high fidelity state preparation, makes them a powerful platform for quantum simulation and information processing~\cite{Cirac:1995,Haffner:2008,Blatt:2008,Bruzewicz:2019, Porras:2004,Deng:PRA:2005,Friedenauer:2008,Kim:2009,Kim:2010,Kim:2011,Islam:2011,Britton:2012,Bermudez:2011,Monroe:2019}. A further enhancement of this setup can be achieved in trapped Rydberg ions, where each ion can be individually excited to a high-lying Rydberg level~\cite{Gallagher:1988,Gallagher:2005, Saffman:2010, Low:2012,Muller:2008,Schmidt-Kaler:2011,Higgins:2017,Higgins:2017PRL,Mokhberi:2019}. The strong dipole-dipole interactions and the interplay between electronic and vibrational degrees of freedom characterizing this system can be exploited to simulate equilibrium and non-equilibrium quantum many-body spin models~\cite{Hague:2012,Nath:2015,Gambetta:2019}, to devise non-classical motional states~\cite{Li:2012}, and for quantum information processing beyond the scalability limitations of conventional ion settings~\cite{Li:2014, Vogel:2019, Zhang:2020}.

 \begin{figure}
	\centering
	\includegraphics[width=\columnwidth]{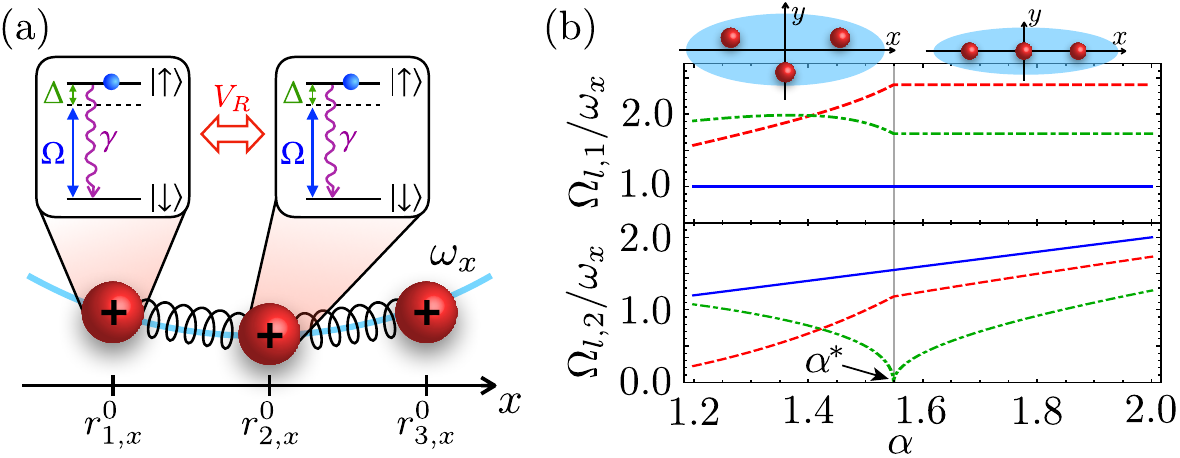}	\caption{{\bf Setup and phonon eigenfrequencies of a three-ion crystal. } (a) Ions are modeled as two-level systems whose ground state, $ \ket{\downarrow} $, is coupled to a Rydberg state, $ \ket{\uparrow} $, by a laser with Rabi frequency $ \Omega $ and detuning $ \Delta $. The state $ \ket{\uparrow} $ spontaneously decays to $ \ket{\downarrow} $ with rate $ \gamma $. The equilibrium positions of the ions $ \bm{r}^0_{n} $ are determined by the interplay between Coulomb repulsion and harmonic confinement. Ions $ m $ and $ n $ interact through the interaction potential $ V_R(\bm{r}_n,\bm{r}_m) $ when both are excited to Rydberg states. (b) Eigenfrequencies $ \Omega_{l,\lambda} $ as a function of the trap aspect ratio $ \alpha=\omega_y/\omega_x $ ($ l=1 $ blue solid, $ l=2 $ red dashed, $ l=3 $ green dash-dotted). For $ \alpha>\alpha^* $, top panel ($ \lambda=1 $) corresponds to longitudinal phonon modes while bottom panel ($ \lambda=2 $) displays the transverse ones. The gray line highlights the structural phase transition at $ \alpha^*=\sqrt{12/5} $. The two configurations of a three-ion chain are sketched on the top. Shaded blue areas represent the trapping region in the $ x-y $ plane. \label{fig:3ionseigenfreq}}
\end{figure}

In this work we demonstrate that the unique intertwining between intrinsically collective vibrational motion and dipole-dipole interactions characterizing trapped Rydberg ions provides a mechanism to engineer long-range and multi-body interactions in state-of-the-art setups. With respect to their neutral counterparts, Rydberg ions offer important experimental advantages. They can be conveniently trapped via state-independent electric potentials and, thus, do not require magic trapping conditions~\cite{Zhang:2011, Barredo:2020, Gambetta:2020}, while the enhanced control over both electronic and vibrational degrees of freedom allows to manipulate their state with an unprecedented degree of fidelity. We investigate the emergence of multi-body interactions by focusing on a chain of $ {}^{88}\mathrm{Sr}^+ $ Rydberg ions confined by harmonic potentials [see Fig.~\ref{fig:3ionseigenfreq}(a)], which has been recently experimentally realized~\cite{Zhang:2020}. We demonstrate that their strength is significantly enhanced in the presence of soft vibrational modes. These occur, e.g., at the linear-to-zigzag transition in a chain of few ions~\cite{Birkl:1992,Raizen:1992,Dubin:1993,Schiffer:1993,Enzer:2000,Morigi:PRE:2004, Fishman:PRB:2008} and also in long linear chains. Here, interactions induced by spin-phonon coupling give rise to non-trivial many-body phenomena, such as a three-body anti-blockade effect. The novel capabilities we unveil in our work show that trapped Rydberg ions are a powerful platform for quantum simulation, allowing for the study of exotic kinetically constrained dynamics~\cite{Sous:2019,Mazza:2020}, long-lived quantum information storage~\cite{Belyansky:2019} and correlated quantum states of matter~\cite{Moessner:2001,Kitaev:2006, Schmidt:2008,Pachos:2004,Capogrosso:2009, Buchler:2007, Bonnes:2010, Will:2010}.
	 

\textit{Spin-phonon coupling induced multi-body interactions.---} 
We consider a quasi one-dimensional chain of $ N $ two-level Rydberg ions confined by a harmonic potential, as sketched in Fig.~\ref{fig:3ionseigenfreq}(a). 
Each ion is modeled as a two-level system (with $ \ket{\downarrow} $ and $ \ket{\uparrow} $ denoting ground and Rydberg state, respectively). The two levels are coupled by a laser field with Rabi frequency $ \Omega $ and detuning $ \Delta $. 
The overall Hamiltonian is
\begin{equation}\label{eq:Hfull}
H=H_\mathrm{ions}+H_\mathrm{L}+H_\mathrm{int},
\end{equation}
with $ H_\mathrm{ions}=\sum_{l,\lambda} \hbar \Omega_{l,\lambda} (a^\dagger_{l,\lambda} a_{l,\lambda} + 1/2) $ describing the vibrational dynamics of an ion crystal confined in a three-dimensional harmonic potential $ v_\mu(r_{n,\mu})=M\omega_{\mu}^2 r_{n,\mu}^2/2 $. Here, $ \bm{r}_n $ is the $ n- $th ion position, $ M $ its mass, and $ \omega_\mu $ the trapping frequency along direction $ \mu=\{x,y,z\} $~\cite{James:1998,Kielpinski:2000,Fishman:PRB:2008}. Bosonic operators $ a^{(\dagger)}_{l,\lambda} $ are associated with the phonon mode $ (l,\lambda)  $ with eigenfrequency $ \Omega_{l,\lambda} $, where $ \lambda $ labels the phonon branches~\cite{Note1}. We assume $ \omega_{x,y}\ll\omega_z $, so that the motion of the ions is effectively confined to the $ x-y $ plane. In Eq.~\eqref{eq:Hfull}, $ H_\mathrm{L}=\sum_{n=1}^{N}\left[\Omega\sigma^x_{n}+\Delta n_{n}\right] $ describes the laser excitation of ions to the Rydberg state. Here, $ n_n=(1+\sigma^z_n)/2 $ and $ \sigma^{\mu}_n $ are the Rydberg number operator and Pauli matrices acting on the $ n- $th ion, respectively. The electrostatic dipole-dipole interaction between pairs of Rydberg excited ions is modeled by $ H_\mathrm{int}=\sum^{\prime}_{m,n} V_R(\bm{r}_m,\bm{r}_n)n_{m} n_{n} $, where $ V_R(\bm{r}_m,\bm{r}_n)=V_R(|\bm{r}_m-\bm{r}_n|) $ and the prime denotes $ m\neq n $. Under typical experimental conditions, the displacements of ions from their equilibrium positions $ \bm{r}_n^0 $ are much smaller than inter-ion distances. Hence, we expand the two-ion potential $ V_R(\bm{r}_m,\bm{r}_n) $ to first order around $ \bm{r}_n^0 $~\cite{Gambetta:2020,Note1}. By substituting the expansion into Eq.~\eqref{eq:Hfull} and performing a polaron transformation to approximately decouple electronic and vibrational degrees of freedom~\cite{Note1,Porras:2004,Deng:PRA:2005,Gambetta:2020}, the full Hamiltonian becomes
\begin{equation}\label{eq:Hprimefinite}
H' \simeq H_\mathrm{ions} + H_\mathrm{spin} + H_\mathrm{res}.
\end{equation}
The spin Hamiltonian $ H_\mathrm{spin}=H_L+H^0_\mathrm{int}+H_\mathrm{int}^\mathrm{eff} $ contains, in addition to the bare dipole-dipole interaction term $ H^0_\mathrm{int}=\sum_{m,n}^\prime V_R(\bm{r}^0_m,\bm{r}^0_n) n_m n_n $, also an \emph{effective} Rydberg interaction contribution
 \begin{equation}\label{eq:Hintefffinite}
H_\mathrm{int}^\mathrm{eff}=-\primesum_{m,n}\primesum_{i,j} \tilde{V}_{mnij} n_m n_n n_i n_j,
\end{equation}
generated as a consequence of the polaron transformation. The latter consists of long-range and multi-body interactions coupling two, three, and four spins. Their strength is encoded in the interaction coefficients
 \begin{equation}\label{eq:Vmnijfinite}
 \tilde{V}_{mnij}=\frac{2} {M}G_{mn}G_{ij}\sum_{\mu,\nu}F^{\mu\nu}_{mi}\bar{R}^0_{mn;\mu}\bar{R}^0_{ij;\nu},
 \end{equation}
where we defined $ \nabla_{r_{m,\mu}}V_R(\bm{r}_m,\bm{r}^0_n)|_{\bm{r}_m^0}\equiv G_{mn}\bar{\bm{R}}^0_{mn} $, with coefficients $ G_{mn} $ describing the magnitude of the gradient of the Rydberg potential and factors $ \bar{\bm{R}}^0_{mn}=(\bm{r}^0_m-\bm{r}^0_n)/|\bm{r}^0_m-\bm{r}^0_n| $ encoding the geometry of the system.  
Effective interactions explicitly depend on the trapping regime via the coupling parameters $ F^{\mu\nu}_{mi}=\sum_{l,\lambda} \Omega_{l,\lambda}^{-2}\mathcal{M}_{ml}^{\mu\lambda}\mathcal{M}_{il}^{\nu\lambda} $, where the normal mode matrices $ \mathcal{M}^{\mu\lambda}_{ml} $ relate local ion displacements to chain normal modes~\cite{Note1}. Hence, the coefficients $ \tilde{V}_{mnij} $ can be controlled by both the Rydberg interaction potential, through its gradient coefficients $ G_{mn} $, and the vibrational structure of the chain. Finally, in Eq.~\eqref{eq:Hprimefinite}, $ H_\mathrm{res} $ contains a residual spin-phonon interaction~\cite{Note1,Porras:2004,Deng:PRA:2005}. In the strong interaction regime, which will be the focus of next sections, its contribution to the spin dynamics is negligible when $ \Omega\ll\Omega^*\sim\min(\Omega_{l,\lambda}^{-1/2}) $~\cite{Note1}. In this case, the electronic and vibrational degrees of freedom decouple.

\begin{figure}
	\centering
	\includegraphics[width=\columnwidth]{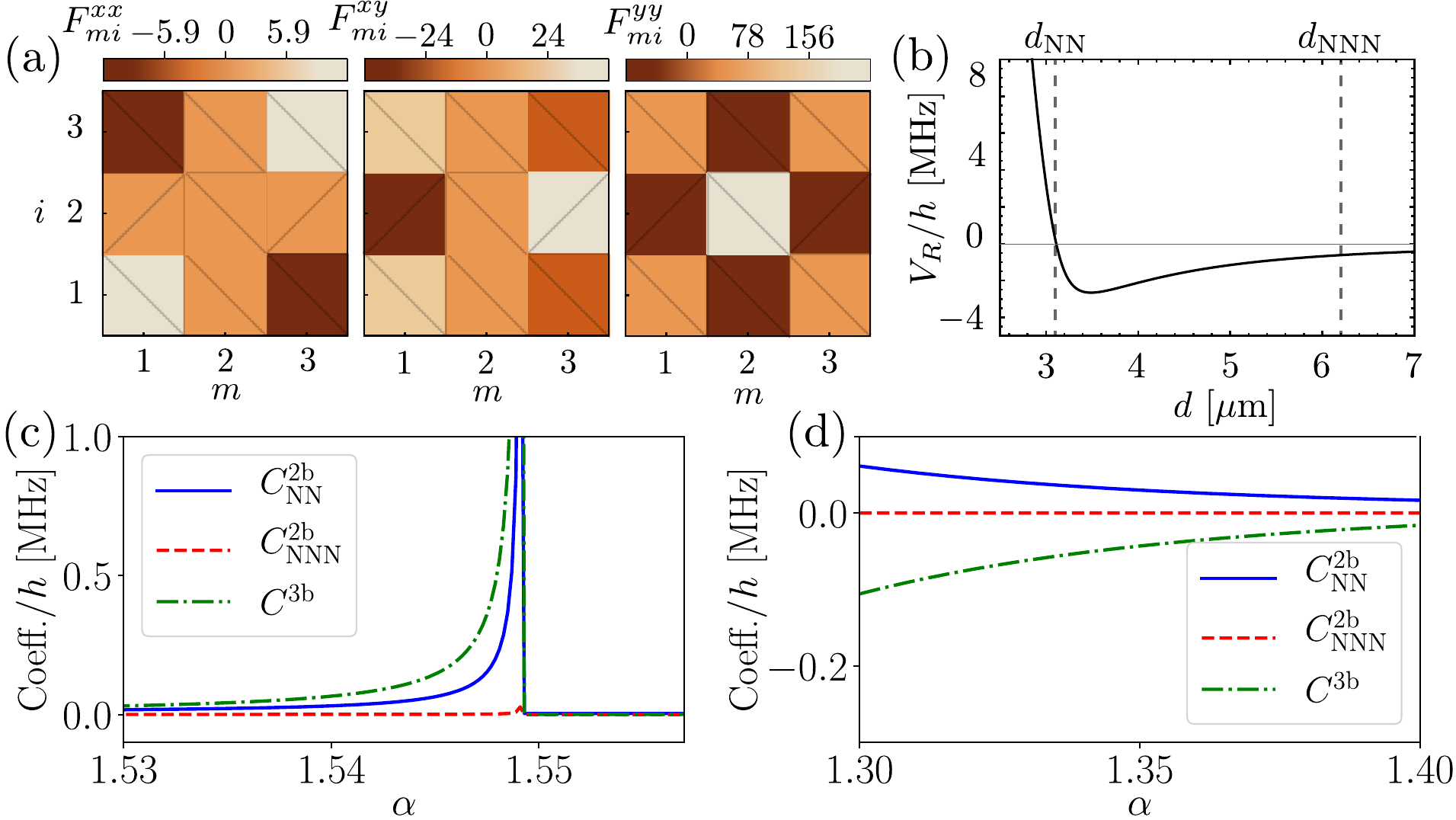}
	\caption{{\bf Effective multi-body interaction in a three-ion chain.} 
		(a) Coupling parameters $ F^{\mu\nu}_{mi} $ (units $ \omega_x^{-2} $) for $ \alpha=1.54 $. 
		(b) Two-ion MW dressed potential $ V_R(d) $ as a function of the ion separation $ d $~\cite{Note1}. Dashed gray lines denote the distance between NNs, $ d_{\mathrm{NN}}=3.1\ \mu\mathrm{m} $, and NNNs, $ d_{\mathrm{NNN}}=2d_{\mathrm{NN}} $, corresponding to a linear chain configuration with longitudinal trapping frequency $ \omega_x=2\pi\times 1.3 $ MHz. 
		Here, $ V_\mathrm{NN}/h=0.25 $ MHz and $ V_\mathrm{NNN}/h=-0.6 $ MHz, with corresponding gradients $ G_\mathrm{NN}/h=23.6\ \mathrm{MHz}/\mu\mathrm{m} $ and $ G_\mathrm{NNN}/h=0.3\ \mathrm{MHz}/\mu\mathrm{m} $.
		(c, d) Effective multi-body interaction coefficients associated with the potential in panel (b) as a function of the trap aspect ratio $ \alpha $. (c) The emergence of a soft mode at the linear-to-zigzag transition leads to a significant enhancement of the effective interactions. 
		(d) For smaller values of $ \alpha $, $ C^{\mathrm{2b}}_{\mathrm{NN(N)}} $ and $ C^{\mathrm{3b}} $ have opposite sign, leading to a competition between attractive and repulsive interactions.
		\label{fig:Coeff3ion}}
\end{figure}

\textit{Three-ion chain.---} 
We first investigate the onset of effective multi-body interactions in a minimal setting of three ions. In this case $ H_\mathrm{int}^{\mathrm{eff}} $ can be explicitly written as
\begin{equation}\label{eq:Hinteff3ions}
H_\mathrm{int}^{\mathrm{eff}}=-C_\mathrm{NN}^\mathrm{2b}(n_1 n_2+n_2 n_3) - C_\mathrm{NNN}^\mathrm{2b}n_1 n_3-C^\mathrm{3b} n_1 n_2 n_3.
\end{equation}
Here, $ C^{\mathrm{2b}}_{\mathrm{NN}} $ and $ C^{\mathrm{2b}}_{\mathrm{NNN}} $ parameterize effective two-body interactions between nearest neighbors (NNs) and next-to-nearest neighbors (NNNs), respectively, while $ C^{\mathrm{3b}} $ describes the three-body contribution. 
Their behavior in the various trapping regimes can be inferred from Eq.~\eqref{eq:Vmnijfinite}. 
A three-ion chain features a second-order phase transition to a zigzag configuration at the critical value $ \alpha^*=\sqrt{12/5}  $ of the trap aspect ratio $ \alpha=\omega_y/\omega_x $~\cite{Fishman:PRB:2008,Note1}. The transition is signaled by the emergence of a \emph{soft mode} with eigenfrequency $ \Omega_{3,2} \rightarrow 0 $ [see Fig.~\ref{fig:3ionseigenfreq}(b)] which, depending on the configuration of the ions, may strongly affect the effective interactions. In the linear regime ($ \alpha>\alpha^* $) the longitudinal and transverse modes of the chain are not coupled and the normal mode matrices $ \mathcal{M}^{\mu\lambda}_{ml} $ vanish when $ \mu\neq\lambda $~\cite{Note1}. In this case, the soft mode is purely transverse and, since $ \bar{R}_{mn;y}\propto r^0_{m,y}-r^0_{m,y}=0 $, it does not contribute to $ \tilde{V}_{mnij} $. This traces back to the fact that transverse displacements hardly affect inter-ion distances and only generate higher-order terms in the expansion of the Rydberg interaction potential. Thus, in a linear chain only longitudinal modes contribute to $ \tilde{V}_{mnij} $ via the coupling parameter $ F^{xx}_{mi} $. As show in Fig.~\ref{fig:3ionseigenfreq}(b), the latter are constant as a function of $ \alpha $ and only depends on $ \omega_x $ as $ F^{xx}_{mi} \propto \Omega_{l,1}\sim\omega_x^{-2}$. We note that, for $ \alpha>\alpha^* $, $ \omega_x $ uniquely determines the distances between ions, which scale as $ \omega_x^{-2/3} $~\cite{Note1}. Since for dipole-dipole interaction potentials larger distances result in smaller gradients, in a three-ion chain it is not possible to arbitrarily “soften” the longitudinal modes and, thus, the magnitude of the effective interactions that can be achieved is strongly limited. In the zigzag configuration ($ \alpha<\alpha^* $), on the contrary, all normal modes possess both longitudinal and transverse components, i.e.~$ \mathcal{M}_{ml}^{\mu\lambda}\neq 0$, $ \forall \mu,\lambda $~\cite{Note1}. As shown in Fig.~\ref{fig:Coeff3ion}(a), the collective and intertwined nature of phonon modes results in non-vanishing coupling parameters $ F^{\mu\nu}_{mi} $, $ \forall \mu,\nu $.  Moreover, $ \bar{R}_{mn;y}\neq0 $ and hence all the couplings $ F^{\mu\nu}_{mi} $ contribute to Eq.~\eqref{eq:Vmnijfinite}. Crucially, due to the presence of the soft mode $ (3,2) $, for $ \alpha\rightarrow(\alpha^*)^- $ one finds $ F^{\mu\nu}_{mi} \sim \Omega_{3,2}^{-2} $, $ \forall \mu,\nu $. We therefore expect that the presence of such a soft mode results in an increase of the effective interaction strength.

Looking at Eq.~\eqref{eq:Vmnijfinite}, one notices that large Rydberg potential gradients $ G_{mn} $ are essential to maximize effects of the effective multi-body interactions. Unfortunately, in trapped Rydberg ions van der Waals interactions are generally weaker than their neutral counterparts and do not give access to large gradients. To overcome this issue, we exploit the interplay between dipole-dipole interactions of microwave (MW) dressed states~\cite{Muller:2008,Li:2014} and MW-tuned F\"orster resonance in a setup of $ {}^{88}\mathrm{Sr}^+ $ trapped Rydberg ions~\cite{Higgins:2017,Higgins:2017PRL,Zhang:2020}. This allows us to obtain the ion-ion interaction potential shown in Fig.~\ref{fig:Coeff3ion}(b)~\cite{Note1}.
The corresponding effective interaction coefficients are shown in Fig.~\ref{fig:Coeff3ion}(c,d) as a function of the trap aspect ratio $ \alpha $.
For $ \alpha>\alpha^* $ their values are fixed and small, as expect from the discussion above. On the other hand, for $ \alpha\rightarrow(\alpha^{*})^{-} $, the two-dimensional configuration of the chain, the mixing between longitudinal and transversal modes, and the emergence of a soft mode result in a significant enhancement of the effective interaction strength. The sign of interaction coefficients is determined by the gradient of the Rydberg potential at NNs and NNNs, $ G_{\mathrm{NN}} $ and $ G_{\mathrm{NNN}} $, respectively. The potential chosen in Fig.~\ref{fig:Coeff3ion}(b) gives $ G_{\mathrm{NN}}<0 $ and $ G_{\mathrm{NNN}}>0 $. Close to the linear-to-zigzag transition, this results in $ C^{\mathrm{2b}}_{\mathrm{NN}}, C^{\mathrm{2b}}_{\mathrm{NNN}}, C^{\mathrm{3b}}>0 $ [see Fig.~\ref{fig:Coeff3ion}(c)], corresponding to attractive interactions. Interestingly, for $ \alpha\lesssim 1.5 $, $ C^{\mathrm{2b}}_{\mathrm{NN}}>0, C^{\mathrm{2b}}_{\mathrm{NNN}}\simeq0$, and $C^{\mathrm{3b}}<0 $, implying a competition between attractive two-body and repulsive three-body effective interactions [see Fig.~\ref{fig:Coeff3ion}(d)]. 

\begin{figure}
	\centering
	\includegraphics[width=\columnwidth]{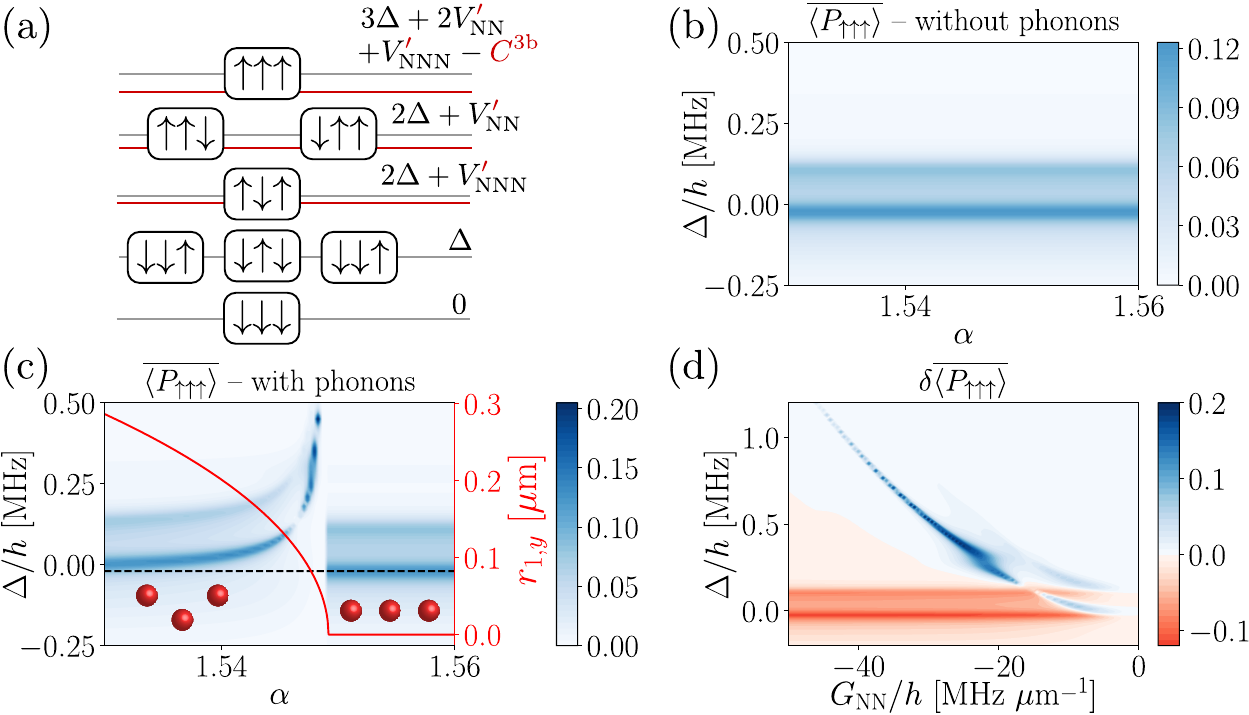}
	\caption{{\bf Three-body spectroscopy.} (a) Electronic levels of a three-ion chain. Here, $ V'_\mathrm{NN}=V_\mathrm{NN}-C^{\mathrm{2b}}_\mathrm{NN} $ and $ V'_\mathrm{NNN}=V_\mathrm{NNN}-C^{\mathrm{2b}}_\mathrm{NNN} $. Red (gray) lines correspond to energy levels in the presence (absence) of effective interactions in the regime of Fig.~\ref{fig:Coeff3ion}(c). (b, c) Time-integrated expectation value of $  P_{\uparrow\uparrow\uparrow} $ as a function of trap aspect ratio $ \alpha $ and detuning $ \Delta/h $ for a system of $ N=3 $ ions without (b) and with (c) effective interactions. 
	In both panels, the initial state is $ \ket{\downarrow\downarrow\downarrow} $ and $ \Omega/h=0.1 $ MHz. Other parameters as in Fig.~\ref{fig:Coeff3ion}. Time averages are evaluated over a $ 50\ \mu\mathrm{s} $ window and we included the finite Rydberg lifetime $ \tau=\gamma^{-1}=30\ \mu\mathrm{s} $ [see Fig.~\ref{fig:3ionseigenfreq}(a)]. In (c), the red solid curve shows the transverse displacement $ r_{1,y}=r_{3,y} $ as a function of $ \alpha $~\cite{Note1}, while the black dashed line highlights the position of the maximum of $  \overline{\av{P_{\uparrow\uparrow\uparrow}}} $ from panel (b). (d) Difference between the time-averaged expectation values of $  P_{\uparrow\uparrow\uparrow} $ with and without effective interactions as a function of $ G_\mathrm{NN}/h $ and $ \Delta/h $ for $ \alpha=1.548 $. Here, $ G_\mathrm{NNN}=0.01 G_\mathrm{NNN} $. Other parameters as in (b, c). \label{fig:facilitation}}
\end{figure}

We now investigate an interaction induced three-body Rydberg anti-blockade regime, a generalization of the well-studied facilitation mechanism in the presence of two-body Rydberg interactions~\cite{Ates:2007,Lesanovsky:2014,Valado:2016,Marcuzzi:2017,Ostmann:2019}. By denoting with $ V_\mathrm{NN} $ and $ V_\mathrm{NNN} $ the bare Rydberg interactions between NNs and NNNs contained in $ H_\mathrm{int}^0 $, respectively, this regime is achieved when [see the level structure in Fig.~\ref{fig:facilitation}(a)] 
\begin{equation}\label{eq:antiblockade}
3\Delta+2(V_{\mathrm{NN}}-C^\mathrm{2b}_{\mathrm{NN}})+(V_{\mathrm{NNN}}-C^\mathrm{2b}_{\mathrm{NNN}})-C^{\mathrm{3b}}=0.
\end{equation}
If ions are prepared in state $ \ket{\downarrow\downarrow\downarrow} $ at time $ t=0 $, an enhancement in the projector on state $ \ket{\uparrow\uparrow\uparrow} $ at subsequent times, $ P_{\uparrow\uparrow\uparrow}(t) $, is expected for values of $ \Delta $ satisfying Eq.~\eqref{eq:antiblockade}. The behavior of the time-integrated expectation value of $ P_{\uparrow\uparrow\uparrow}(t) $, denoted by $ \overline{\langle P_{\uparrow\uparrow\uparrow}\rangle} $, is shown in Fig.~\ref{fig:facilitation}(b, c). Panel (b) shows the case with bare Rydberg interactions only (i.e., with $ C^{\mathrm{2b}}_{\mathrm{NN}}=C^{\mathrm{2b}}_{\mathrm{NNN}}=C^{\mathrm{3b}}=0 $), while effects of multi-body interactions are displayed in panel (c). For $ \alpha\rightarrow (\alpha^*)^- $, effective interactions modify significantly the value of $ \Delta $ satisfying Eq.~\eqref{eq:antiblockade}. This results in a shift of the peak of $ \overline{\langle P_{\uparrow\uparrow\uparrow}\rangle} $. As can be seen in Fig.~\ref{fig:facilitation}(d), showing the difference $ \delta\overline{\langle P_{\uparrow\uparrow\uparrow}\rangle} $ between the time-integrated expectation values of $ P_{\uparrow\uparrow\uparrow}(t) $ with and without the contributions of $ H_\mathrm{int}^{\mathrm{eff}} $, the presence of phonon-mediated multi-body interactions leads to a clear spectroscopic signature.

The latter provides a sensitive tool to locate the critical point of the linear-to-zigzag transition, allowing for a significant improvement over state-of-the-art methods~\cite{Enzer:2000,Gong:2010,Zhang:2017}. Indeed, the typical resolution of current direct camera images of the ions is $ \approx0.5\mu\mathrm{m} $ and, being the transition a second order one, they can hardly reveal the small ion displacements along the traverse direction for $\alpha\approx\alpha^* $. In contrast, from Fig.~\ref{fig:facilitation}(c) we see that close to the critical point a traverse displacement of $ \approx 0.1\ \mu\mathrm{m}$ corresponds to a shift of $ \approx0.2\ \mathrm{MHz} $ in the peak of $ \overline{\langle P_{\uparrow\uparrow\uparrow}\rangle} $, which can be easily detected via spectroscopic measurements~\cite{Note2}.

\textit{Infinite linear chain.---} 
In the previous section we showed that in a three-ion chain spin-phonon coupling induced interactions are strongly amplified by the emergence of a soft mode. We now inspect the behavior of $ H_\mathrm{int}^{\mathrm{eff}} $ in longer chains, where system properties explicitly depend on the number of ions, $ N $~\cite{Dubin:1993,Schiffer:1993,Yan:2016}.
In particular, by increasing $N$, one can decrease the inter-ion distance in the central region of the chain even in the presence of a weak longitudinal confinement. This fact can be exploited to engineer soft modes even in the linear configuration and, hence, it allows to overcome the restrictions on the strength of effective interactions we discussed for a linear three-ion chain. As we will show, the increased flexibility provided by a denser vibrational spectrum also provides a convenient handle to control the range of the effective interactions.

To gain insights into the phenomenology of this case, we consider the infinite chain limit $ N\rightarrow\infty $, which provides a good description of the central region of long yet finite chains~\cite{Morigi:PRL:2004, Morigi:PRE:2004}. In the linear regime~\cite{Fishman:PRB:2008}, the equilibrium positions of the ions are $ \bm{r}^0_{n}=(n d, 0) $, with $ d $ being the fixed inter-ion distance and $ n\in\{0,\pm 1,..., \pm \infty\} $. To mimic the effect of a longitudinal confinement, we replace the harmonic trapping potential along the $ x $ axis with a periodic one commensurate with the lattice spacing, $ v_x(r_{n,x})= - M \omega_x (d/2\pi)^2 \cos(2\pi r^0_{n,x}/d) $.  Expanding ions' coordinates in Fourier modes and generalizing the steps leading to Eq.~\eqref{eq:Hprimefinite}, we obtain a Hamiltonian $ H' $ with the same form as Eq.~\eqref{eq:Hprimefinite}~\cite{Note1}. In analogy with the three-ion case, in the linear configuration longitudinal and transverse modes are not coupled (i.e., $ F^{\mu\nu}_{mi}=0 $ for $ \mu\neq\nu $) and $ \bar{R}^0_{mn;y}=0 $. Thus, only longitudinal modes contribute to $ \tilde{V}_{mnij} $ via $ F^{xx}_{mi}\equiv F(s)=(2\pi)^{-1}\int_{-\pi}^{\pi} \mathrm{d}k\ [\Omega_{x}(k)]^{-2} e^{-i k s} $, with $ s=m-i $, $ -\pi \leq k <\pi $ defining the wavevector of the first Brillouin zone, and $ \Omega_{x}(k) $ the eigenfrequency of phonon mode $ (k,x) $~\cite{Note1}.

\begin{figure}
	\centering
	\includegraphics[width=\columnwidth]{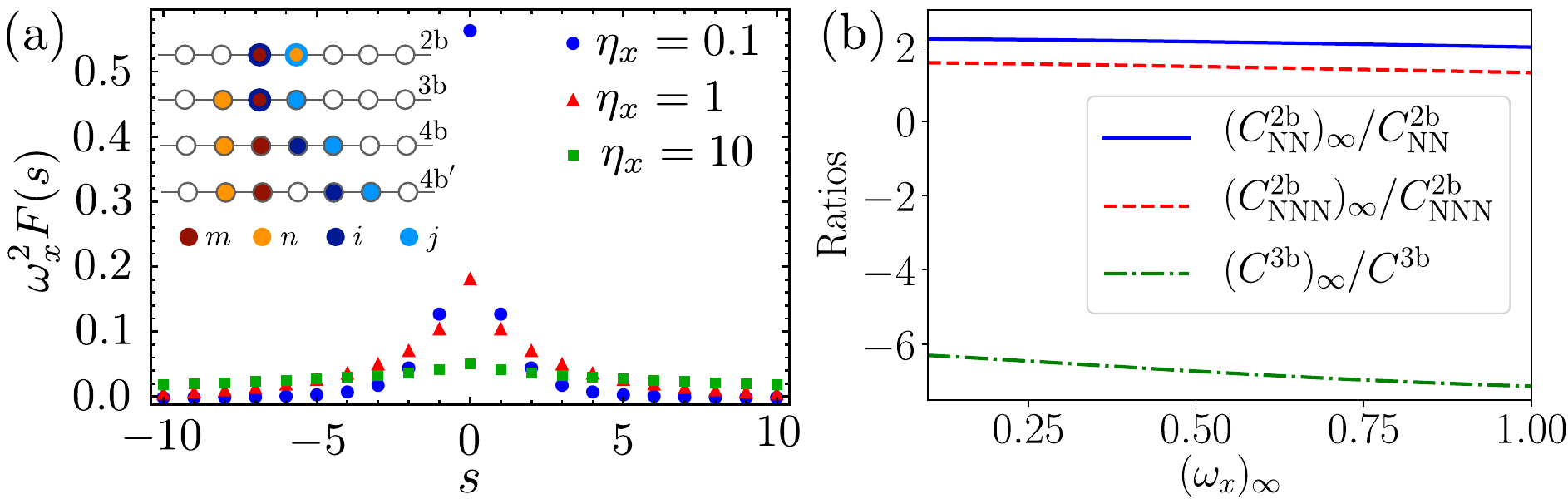}
	\caption{{\bf Effective multi-body interactions in an infinite linear chain.} (a) Coupling parameters $ \omega_x^2 F (s)$ as a function of $ s=m-i $ for $ \eta_x=0.1 $ (stiff regime), $ \eta_x=1 $ (intermediate regime), $ \eta_x=10 $ (soft regime). Inset: Examples of possible contributions to $ H_\mathrm{int}^{\mathrm{eff}} $. Since they are all $ \propto F(s=m-i) $, in the stiff limit the $ \mathrm{4b}^\prime $ term is strongly suppressed with respect to the 4b one. (b) Ratios between the effective interaction coefficients in a linear infinite chain and the corresponding ones for the three-ion case as a function of the infinite chain trapping frequency, $ (\omega_x)_\infty $ (units $ 2\pi\times\mathrm{MHz} $). Distance between NNs is fixed as $ d=3.1\ \mu\mathrm{m} $ (corresponding to $ \omega_x=2\pi\times1.3 $ MHz for the three-ion chain). Rydberg potential parameters are as in Fig.~\ref{fig:Coeff3ion}, while $ \eta_x>1 $ and $ N>4$ throughout the whole range of $ (\omega_x)_\infty $ considered.  \label{fig:Fmi}}
\end{figure}

The dense vibrational spectrum leads to two different regimes for the ion dynamics which can be controlled via the trap parameter $ \eta_x= V_0/(Md^3\omega^2_x) $~\cite{Deng:PRA:2005} and which allows to substantially modify the behavior of $ F(s)$ [see Fig.~\ref{fig:Fmi}(a)]. For $ \eta_x\ll 1 $ (stiff limit), ions behave as independent harmonic oscillators, while for $ \eta_x\gg1  $ (soft limit) phonon modes describe genuinely collective excitations. In the stiff limit, $ F(s)$ is peaked around $ s=0 $, implying that dominant contributions to $ H^\mathrm{eff}_{\mathrm{int}} $ consist of connected strings of neighboring two-, three-, and four-body terms. On the other hand, for $ \eta_x\gtrsim 1$, $ F(s)$ has a broader distribution and is non-negligible also for $ |s|>0 $. As a consequence, exotic long-range and multi-body interaction terms arise in $ H^\mathrm{eff}_{\mathrm{int}} $, as shown, e.g., in the bottom row of the inset in Fig.~\ref{fig:Fmi}(a). This broad spectrum of possible interaction patterns, allowed by the collective nature of phonon modes and the precise control over the chain trapping parameters, is in contrast with the case of Rydberg atom tweezer arrays, where only short-range two- and three-body interactions can be engineered~\cite{Gambetta:2020}.


\textit{Three-body spectroscopy of a long chain.---} 
The previous discussion allows to gain an understanding of the three-body spectroscopy of a long yet finite chain, which can be experimentally investigated in trapped Rydberg ion simulators. Indeed, in a long enough chain, long wavelength soft modes, which give the largest contribution to $ H_\mathrm{int}^{\mathrm{eff}} $, coincide with good approximation with the ones of the corresponding infinite chain limit~\cite{Morigi:PRE:2004,Morigi:PRL:2004}. Moreover, due to the finite lifetime of Rydberg excitations $ \tau=\gamma^{-1} $, with $ \gamma $ the spontaneous decay rate [see Fig.~\ref{fig:3ionseigenfreq}(a)], the vibrational spectrum of the chain can be considered as continuous when energy gaps between phonon modes are smaller than $ \gamma $. For $ {}^{88}\mathrm{Sr}^+ $ ions (with $ \tau\approx30\ \mu\mathrm{s} $) in a trap with $ \omega_x=2\pi\times0.2 $ MHz, the above conditions are both met for chains with $ N\gtrsim20 $, which are within the reach of current state-of-the-art setups~\cite{Zhang:2020}.
The possibility to control inter-ion distances in the central region of a long chain by tuning $ N $ allows to employ a weaker longitudinal confinement, which results in the emergence of soft modes $ \Omega_x(k) \sim \omega_x $ near $ k\approx 0 $. Thus, an enhancement of the effective interactions can be obtained even in the linear configuration. To quantify this effect, in Fig.~\ref{fig:Fmi}(b) we plot the ratio between effective two and three-body coefficients for an infinite chain and the corresponding ones for the three-ion setup shown in Fig.~\ref{fig:Coeff3ion}(c). To make the comparison meaningful, we fix the inter-ion distance in the infinite chain as $ d=d_\mathrm{NN} $, with $ d_\mathrm{NN} $ given in Fig.~\ref{fig:Coeff3ion}(c). When  the infinite chain trapping frequency, $ (\omega_x)_\infty $, is varied,  $ d $ can be kept fixed by adjusting $ N $. In principle, by adding more ions to the chain while keeping $ (\omega_x)_\infty $ constant, smaller values of $ d $ can be accessed, leading to stronger effective interactions.
Looking at Fig.~\ref{fig:Fmi}(b), we also note that in an infinite chain $ C^{\mathrm{2b}}_{\mathrm{NN(N)}} $ and $ C^{\mathrm{3b}} $ have opposite sign. This fact, due to the different vibrational structure, allows to investigate the competition between attractive and repulsive effective interactions. We therefore expect that trapped Rydberg ions will give access to different interaction regimes, paving the way to the study of quantum magnetism and frustration phenomena in the presence of exotic multi-body effects.

\textit{Conclusions.---} 
We investigated the emergence of long-range multi-body interactions in a chain of trapped Rydberg ions induced by the coupling between phonon modes and ion-ion Rydberg interactions. We showed that these interactions are extremely sensitive to the chain equilibrium configuration and vibrational regimes, such as the emergence of soft modes. By employing realistic parameters from a state-of-the-art setup of ${}^{88}\mathrm{Sr}^+$ Rydberg ions, we demonstrated that they result in a neat signature of the linear-to-zigzag transition in ion Rydberg state spectroscopic signal. The intertwining between chain configuration, vibrational structure, and effective interactions illustrated in this work provides a versatile mechanism to investigate quantum dynamics in the presence of non-trivial multi-body interactions and exotic constraints. 


\begin{acknowledgments}	
	The research leading to these results has received funding from the EPSRC Grant No. EP/M014266/1, the EPSRC Grant No. EP/R04340X/1 via the QuantERA project “ERyQSenS”, and the Deutsche Forschungsgemeinschaft (DFG) within the SPP 1929 Giant interactions in Rydberg Systems (GiRyd). CZ and MH acknowledge support from the Swedish Research Council (TRIQS), the QuantERA ERA-NET Cofund in Quantum Technologies (ERyQSenS), and the Knut \& Alice Wallenberg Foundation (WACQT). CZ acknowledges the hospitality of the University of Nottingham. IL acknowledges funding from  the  “Wissenschaftler R\"uckkehrprogramm GSO/CZS” of the Carl-Zeiss-Stiftung  and  the  German  Scholars  Organization e.V.,  as  well as through The Leverhulme Trust [Grant No. RPG-2018-181]. WL acknowledges funding from the UKIERI-UGC Thematic Partnership No. IND/CONT/G/16-17/73, and the Royal Society through the International Exchanges Cost Share award No. IEC$\backslash$NSFC$\backslash$181078.
\end{acknowledgments}	 

\footnotetext[1]{See Supplemental Material, which contains Ref.~\cite{Solyom:vol1:2007}.}
\footnotetext[2]{The divergence of the shift for $ \alpha\rightarrow(\alpha^*)^- $ is an artifact caused by the divergence of the characteristic length associated with the soft mode, $ \ell_\mathrm{s} $, which leads to the breakdown of the small displacement approximation [see Eq.~(S14) of Ref.~\cite{Note1}]. For $ r_{1,y}\approx0.1 \mu\mathrm{m} $ one finds $ \ell_\mathrm{soft}/d_\mathrm{NN}\approx 0.01 $, which is well inside the validity region of the expansion. }


\bibliography{bibliography.bib} 

\pagebreak

\widetext
\begin{center}
	\textbf{\large Supplemental Material for "Long-range multi-body interactions and three-body anti-blockade in a trapped Rydberg ion chain"}
\end{center}

\setcounter{equation}{0}
\setcounter{figure}{0}
\setcounter{table}{0}
\makeatletter
\renewcommand{\theequation}{S\arabic{equation}}
\renewcommand{\thefigure}{S\arabic{figure}}

\renewcommand{\bibnumfmt}[1]{[S#1]}


\par
\begingroup
\leftskip2cm
\rightskip\leftskip
\small In this Supplemental Material we provide additional details on the dynamics of a system of singly charged ions trapped in a three-dimensional harmonic potential. We then comment on the derivation of Eqs.~(2), (3), and (4) of the main text and provide details of the polaron transformation employed to decouple spin and phonon degrees of freedom. Finally, we discuss the microwave dressing scheme leading to the  ion-ion potential shown in Fig.~2(b) of the main text in a state-of-the-art setup of trapped Rydberg ions. 
\par
\endgroup

\section{Dynamics of a trapped ion chain}
In this section we provide additional details on the dynamics of a chain of singly charged ions interacting through Coulomb repulsion and trapped in a three-dimensional harmonic potential. In the spirit of the Born-Oppenheimer approximation, we neglect effects of the Rydberg interactions on the dynamics of the ions. The Hamiltonian of the system is
\begin{equation}\label{SM:eq:Hions}
	H_\mathrm{ions}=T_\mathrm{ions}+V_\mathrm{ions},
\end{equation}
with kinetic energy term $ T_\mathrm{ions}=\sum_{\mu}\sum_n p^2_{n,\mu}/(2M)$, with $ \bm{p}_n $ ions' momenta, and potential energy
\begin{equation}\label{SM:eq:Vions}
	V_\mathrm{ions}=\sum_{n,\mu} v_{\mu}(r_{n,\mu}) + \frac{1}{2}\primesum_{m,n}\frac{V_0}{|\bm{r}_m-\bm{r}_n|}.
\end{equation}
The latter consists of the contributions due to the harmonic trap, with $ v_\mu(r_{n,\mu})=M\omega_{\mu}^2 r_{n,\mu}^2/2 $ (being $ \bm{r}_n $ the position of the $ n- $th ion, $ M $ its mass, and $ \omega_\mu $ the trapping frequency along direction $ \mu=\{x,y,z\}\equiv\{1,2,3\} $), and of the Coulomb repulsion between the ions, with $ V_0=e^2/(4\pi\epsilon_0) $. For the sake of simplicity, we assume $ \omega_{z}\gg\omega_{x,y} $ so that the motion of the ions is confined to the $ x-y $ plane. The equilibrium positions of the ions, $ \bm{r}^0_n $, are determined by solving $ \nabla_{\bm{r}_n}V_\mathrm{ions}|_{\bm{r}_n=\bm{r}_n^0}=0,\, \forall n $. Of particular interest is the case of the three-ion chain considered in main text. Here, from symmetry considerations, one obtains $ r^0_{1,x}=-r^0_{3,x}=X\ell $, $ r^{0}_{2,x}=0 $, $ r^0_{1,y}=-2r^0_{2,y}=r^0_{3,y}=Y\ell $, with $ X,Y>0 $ and $ \ell^3=V_0/(M\omega_x^2) $ a characteristic length scale of the system~\cite{Fishman:PRB:2008,James:1998}. As discussed in the main text, the chain undergoes a linear-to-zigzag phase transition at the critical value of the trap aspect ratio $ \alpha^*=\sqrt{12/5} $, being $ \alpha=\omega_y/\omega_x $. In the linear regime, for $ \alpha<\alpha^*, $ we get $ X=(5/4)^{1/3} $ and $ Y=0 $, while in transverse one, i.e.~when $\alpha<\alpha^*$, $ X=[4(1-\alpha^2/3)]^{-1/3} $ and $ Y=[(3/\alpha^2)^{2/3}-X^2]^{1/2}/3 $~\cite{Fishman:PRB:2008}.

At low temperature, ions perform small oscillations around their equilibrium positions $ \bm{r}^0_n $~\cite{Kielpinski:2000}. In this limit, $ V_\mathrm{ions} $ can be expanded to the second order in the displacements $ \bm{q}_n= \bm{r}_n- \bm{r}^0_n $, leading to
\begin{equation}\label{SM:eq:Vsmallq}
	V_{\mathrm{ions}}=\frac{M}{2}\sum_{m,\mu; n,\nu} \mathcal{K}^{\mu\nu}_{mn} q_{m,\mu} q_{n,\nu},
\end{equation}
with $\mathcal{K}^{\mu\nu}_{mn}=M^{-1}\partial^2 V_{\mathrm{ions}}/(\partial r_{m,\mu}\partial r_{n,\nu})|_{\bm{r}_{m,n}=\bm{r}^0_{m,n}} $ the $ 2N \times 2N $ ions' dynamical matrix~\cite{Solyom:vol1:2007} and $ \mu,\nu=\{x,y\} $.  For the specific case of a three-ion chain one explicitly has
\begin{equation}\label{key}
	\mathcal{K}=\begin{pmatrix}
		\mathcal{K}^{xx} & \mathcal{K}^{xy}\\
		\mathcal{K}^{yx} & \mathcal{K}^{yy}
	\end{pmatrix},
\end{equation}
with 
\begin{subequations}
	\begin{align}\label{key}
		\frac{\mathcal{K}^{xx}}{\omega_x^2}&=\begin{pmatrix}
			1 + a + b_x & -b_x     &-a\\
			-b_x           & 1+2b_x & -b_x\\
			-a               &-b_x       & 1 + a + b_x
		\end{pmatrix},\\ \frac{\mathcal{K}^{yy}}{\omega_x^2}&=\begin{pmatrix}
			\alpha^2 + a/2 + b_y & -b_y     &a/2\\
			-b_y           & \alpha^2+2b_y & -b_y\\
			-a/2               &-b_y       & \alpha^2 + a/2 + b_y
		\end{pmatrix},\\
		\frac{\mathcal{K}^{xy}}{\omega_x^2}&= \frac{\mathcal{K}^{yx}}{\omega_x^2}=\begin{pmatrix}
			-c          & c       &0\\
			c           & 0       & -c\\
			0           &-c       &c
		\end{pmatrix}.
	\end{align}
\end{subequations}
Here, $ a=1/4X^3 $, $ b_x= 3X^2/(X^2+9Y^2)^{5/2} - 1/(X^2+9Y^2)^{3/2} $, $ b_y= 27Y^2/(X^2+9Y^2)^{5/2} - 1/(X^2+9Y^2)^{3/2} $, and $ c= 9XY/ (X^2+9Y^2)^{5/2}$. Note that, in the linear configuration $ Y=0 $ and, therefore, $ \mathcal{K}^{xy}=\mathcal{K}^{yx}=0$, i.e.~the dynamics along $ x $ and $ y $ are decoupled.

Phonon modes and vibrational eigenfrequencies $ \Omega_{m,\mu} $ of the system are obtained by solving the eigensystem associated to $ \mathcal{K} $ in Eq.~\eqref{SM:eq:Vsmallq}, i.e., 
\begin{equation}\label{SM:eq:MKM}
	\mathcal{M}^{-1} \mathcal{K} \mathcal{M} = \mathcal{D}, \quad \text{with}\quad\mathcal{D}^{\mu\nu}_{mn}=\Omega^2_{m,\mu} \delta_{m,n}\delta_{\mu,\nu},
\end{equation}
where $ \delta_{m,n} $ denotes the Kronecker delta and in the normal modes matrix $ \mathcal{M} $ the $ k $-th column corresponds to the (normalized) $ k $-th eigenvector of $ \mathcal{K} $. Considering again the three-ion chain investigated in the main text, the normal normal mode matrix can be written as 
\begin{equation}\label{key}
	\mathcal{M}=\begin{pmatrix}
		\mathcal{M}^{xx} & \mathcal{M}^{xy}\\
		\mathcal{M}^{yx} & \mathcal{M}^{yy}
	\end{pmatrix}.
\end{equation}
In the linear regime the dynamical matrix $ \mathcal{K} $ is block diagonal and, hence, $ \mathcal{K}^{xx} $ and $ \mathcal{K}^{yy} $ can be diagonalized independently of each other. Their normal modes represent purely longitudinal and transverse oscillations, respectively, with $ \mathcal{M}^{xy}_{mn}=\mathcal{M}^{yx}_{mn}=0 $. On the contrary, in the zigzag configuration one has to diagonalize the full dynamical matrix $ \mathcal{K} $ and, in general, $\mathcal{M}^{\mu\nu}\neq0$, $ \forall \mu, \nu $. Therefore, normal modes possess both longitudinal and transverse components and the oscillations along $ x $ and $ y $ are strongly coupled.

The annihilation and creation operators
$ a_{l,\lambda} $ and $ a^\dagger_{l,\lambda} $ of the phonon mode $ (l,\lambda) $, with $ \lambda=\{1,2\} $ labeling the two phonon branches, are then introduced as
\begin{equation}\label{SM:eq:qpfinite}
	q_{m,\mu}=\sum_{n,\nu}\mathcal{M}^{\mu\nu}_{mn} \ell_{n,\nu}(a^\dagger_{n,\nu}+a_{n,\nu})\quad \text{and}\quad 
	p_{m,\mu}=\sum_{n,\nu}\mathcal{M}^{\mu\nu}_{mn} i \wp_{n,\nu} (a^\dagger_{n,\nu}-a_{n,\nu}),
\end{equation}
with $ \ell_{n,\nu}=\sqrt{\hbar/(2M\Omega_{n,\nu})} $ and $ \wp_{n,\nu}=\sqrt{M\hbar\Omega_{n,\nu}/2} $ being the characteristic length and momentum associated with the harmonic trapping potential, respectively. In terms of the phonon mode operators, $ H_{\mathrm{ions}} $ can be written as in Eq. (1) of the main text, i.e.,
\begin{equation}\label{SM:eq:Hionsfinite}
	H_\mathrm{ions}=\sum_{l,\lambda} \hbar \Omega_{l,\lambda} \left(a^\dagger_{l,\lambda} a_{l,\lambda} + \frac{1}{2}\right).
\end{equation}

In the case of an infinite chain, ions' coordinates $ q_{m,\mu} $ and $ p_{m,\mu} $ can be expanded in Fourier modes as
\begin{equation}
	q_{m,\mu}=\frac{1}{\sqrt{2\pi}} \int^{\pi}_{-\pi} \mathrm{d}k\ e^{-i k m} \ell_{\mu}(k) [a^\dagger_\mu(k)+a_\mu(-k)]\quad\text{and}\quad p_{m,\mu}=\frac{1}{\sqrt{2\pi}} \int^{\pi}_{-\pi}\mathrm{d}k\ e^{i k m}i\wp_\mu(k)[a^\dagger_\mu(k)-a_\mu(-k)],
\end{equation}
with $ -\pi \leq k <\pi $ defining the wavevector of the first Brillouin zone, $ \ell_{\mu}(k)=\sqrt{\hbar/[2M\Omega_\mu(k)]} $, and $ \wp_\mu(k)=\sqrt{M\hbar\Omega_{\mu}(k)/2} $. Here, $ a_{\lambda}(k) $ and $ a^\dagger_\lambda(k) $ are  annihilation and creation operators
of the phonon mode $ (k,\lambda) $, respectively. In terms of the latter, the generalization of Eq.~\eqref{SM:eq:Hionsfinite} to an infinite chain is
\begin{equation}\
	H_{\mathrm{ions}}=\sum_{\lambda}\int_{-\pi }^{\pi} \mathrm{d}k\ \left[\hbar \Omega_{\lambda}(k) a^{\dagger}_{\lambda}(k)a_{\lambda}(k)+\frac{1}{2}\right],
\end{equation}
where normal mode eigenfrequencies are given by~\cite{Fishman:PRB:2008}
\begin{equation}\label{SM:eq:eigsinfinite}
	\Omega^2_{\lambda}(k)=\omega^2_{\lambda}\left[1-4c_\lambda\eta_\lambda\sum_{s=1}^{\infty} \frac{1}{s^3}\sin^2 (k s)\right],
\end{equation}
with $ (c_x,c_y)=(-2,1) $ and the parameter $ \eta_\lambda= V_0/(Ma^3\omega^2_\lambda) $  characterizing the trapping regime of the system. 

\section{Effective Hamiltonian and polaron transformation}
Here, we discuss in more details the approximations leading to Eqs.~(2), (3), and (4) of the main text. Starting from Eq.~(1), in the first step we expand the Rydberg interaction potential $ V_R(\bm{r}_m,\bm{r}_n) $ around ions' equilibrium positions $ \bm{r}_n^0 $ to the first order in the displacements $ \bm{q}_n/d=(\bm{r}_n-\bm{r}_n^0)/d $, being $ d $ the typical distance between neighboring ions. We obtain,
\begin{equation}\label{SM:eq:Vapprox}
	V_R(\bm{r}_m,\bm{r}_n)\approx V_R(\bm{r}_m^0,\bm{r}_n^0) + \nabla_{\bm{r}_m}V_R(\bm{r}_m,\bm{r}_n^0)|_{\bm{r}_m=\bm{r}_m^0}\cdot\delta \bm{r}_m + \nabla_{\bm{r}_n}V_R(\bm{r}_m^0,\bm{r}_n)|_{\bm{r}_n=\bm{r}_n^0}\cdot\delta \bm{r}_n.
\end{equation}
Inserting this expansion into the interaction Hamiltonian $ H_\mathrm{int} $ of Eq.~(1) we obtain
\begin{equation}\label{SM:eq:Hintfinite}
	H_\mathrm{int}= \underset{H_{\mathrm{int}}^0}{\underbrace{\primesum_{m,n}V_R(\bm{r}_m^0,\bm{r}_n^0) n_m n_n}}+\sum_{l,\lambda}\primesum_{m,n} W^{\lambda}_{mnl} \left(a^\dagger_{l,\lambda} + a_{l,\lambda}\right) n_m n_n+O(\ell_{\mathrm{min}}^2/d^2),
\end{equation}
where $ \ell_{\mathrm{min}}=\min(\ell_{l, \lambda}) $ is the smallest characteristic length of the harmonic confinement, the normal mode matrices $ \mathcal{M}^{\mu\lambda}_{ml} $ are defined in Eq.~\eqref{SM:eq:MKM}, and $ W^{\lambda}_{mnl}=2 \ell_{l,\lambda} \sum_{\mu} G_{mn}\bar{R}^0_{mn;\mu}\mathcal{M}^{\mu\lambda}_{ml} $, where we introduced $ \nabla_{r_{m,\mu}}V_R(\bm{r}_m,\bm{r}^0_n)|_{\bm{r}_m^0}=G_{mn}\bar{\bm{R}}^0_{mn} $, with $ G_{mn} $ being the magnitude of the gradient of the Rydberg potential and $ \bar{\bm{R}}^0_{mn}=(\bm{r}^0_m-\bm{r}^0_n)/|\bm{r}^0_m-\bm{r}^0_n| $. On the right hand side of Eq.~\eqref{SM:eq:Hintfinite}, the spin-phonon coupling in the second term, whose strength is parameterized by $ W^{\lambda}_{mnl} $, is linear in the phonon annihilation and creation operators. Hence, it can be eliminated by applying the polaron transformation $ U=e^{-S} $~\cite{Porras:2004,Deng:PRA:2005,Gambetta:2020}, defined by
\begin{equation}\label{SM:S}
	S=-\sum_{l,\lambda}\beta^\lambda_{l} \left(a^\dagger_{l,\lambda}-a_{l,\lambda}\right)\quad \text{and}\quad \beta^\lambda_l=\primesum_{m,n}\frac{W^\lambda_{mnl}}{\hbar \Omega_{l,\lambda}} n_m n_n.
\end{equation}
The transformed Hamiltonian reads
\begin{equation}\label{SM:eq:H_CS}
	H'=UHU^{-1} =H_\mathrm{ions}+ \underset{H_{\mathrm{spin}}}{\underbrace{\sum_{n}\left[\Omega\sigma^x_{n}+\Delta n_{n}\right]+H_{\mathrm{int}}^0+H_{\mathrm{int}}^{\mathrm{eff}}}}+H_\mathrm{res}+O(\ell_{\mathrm{min}}^2/d^2).
\end{equation}
Here, $ H_\mathrm{ions} $ is given in Eq.~\eqref{SM:eq:Hionsfinite} and $ H_\mathrm{spin} $ describes the dynamics of a (decoupled) system of spins interacting both through bare Rydberg interactions and effective multi-body interactions described by
\begin{equation}\label{SM:eq:Hintefffinite}
	H_\mathrm{int}^\mathrm{eff}=-\primesum_{m,n}\primesum_{i,j} \tilde{V}_{mnij} n_m n_n n_i n_j,\qquad \text{ with coefficients}\quad \tilde{V}_{mnij}=\sum_{l,\lambda} \frac{W^\lambda_{mnl}W^\lambda_{ijl}}{\hbar \Omega_{l,\lambda}}.
\end{equation}
In Eq.~\eqref{SM:eq:H_CS}, $ H_\mathrm{res} $ represents a residual spin-phonon coupling arising as a consequence of the polaron transformation and is given by~\cite{Porras:2004, Deng:PRA:2005,Gambetta:2020}
\begin{equation}\label{key}
	H_\mathrm{res}=\Omega\left(U\sum_{n}\sigma_n^x U^{-1}-\sum_n\sigma_n^x\right)=i\Omega\primesum_{m,n}\sum_{l,\lambda} \frac{W^\lambda_{mnl}}{\hbar \Omega_{l,\lambda}} \left( n_m \sigma^y_n + \sigma^y_m n_n \right)\left(a^\dagger_{l,\lambda} - a_{l,\lambda} \right).
\end{equation}
In general, the contribution of $ H_\mathrm{res} $ to the system dynamics can be neglected when $ \mathrm{max}[|W^{\lambda}_{mnl}|/(\hbar \Omega_{l,\lambda})]\ll 1 $~\cite{Deng:PRA:2005,Porras:2004}. This condition can be relaxed in the classical limit $ \Omega\rightarrow0 $, where the polaron transformation of Eq.~\eqref{SM:S} results in a perfect spin-phonon decoupling~\cite{Gambetta:2020}. Moreover, in the strong effective interaction regime reached when soft modes emerge, such as, e.g., close to the structural phase transition in the three-ion chain considered in the main text,  effects of $ H_\mathrm{res} $ are negligible with respect to $ H_\mathrm{int}^{\mathrm{eff}} $ when $ \Omega\ll\min(|\hbar \tilde{V}_{mnij}\Omega_{l,\lambda}/W^{\lambda}_{mnl}|) $. In the presence of a single soft mode, with corresponding eigenfrequency $ \Omega_{\mathrm{soft}} \rightarrow 0 $, the latter condition can be further simplified by noticing that, in this case, the dominant contribution to both $ H_\mathrm{res} $ and $ H_\mathrm{int}^{\mathrm{eff}} $ is given by the soft mode only while all the other modes can be neglected. Thus, by denoting with $ a^{(\dagger)}_\mathrm{soft} $ and $ W^\mathrm{soft}_{mn} $ the bosonic operators and the spin-phonon coupling strength associated with the soft mode, respectively, one can write $ \tilde{V}_{mnij}\approx  W^{\mathrm{soft}}_{mn}W^{\mathrm{soft}}_{ij}/(\hbar \Omega_{\mathrm{soft}})  $ and $ H_\mathrm{res}\approx i\Omega\sum^{\prime}_{m,n} W^{\mathrm{soft}}_{mn}/(\hbar \Omega_{\mathrm{soft}}) ( n_m \sigma^y_n + \sigma^y_m n_n) (a^\dagger_{\mathrm{soft}} - a_{\mathrm{soft}} ) $. Assuming $ G_\mathrm{NN}\gg G_\mathrm{NNN} $, we can approximate $ W^{\mathrm{soft}}_{mn}\approx \ell_{\mathrm{soft}} G_\mathrm{NN} $, where we used that $ 2\sum_{\mu} \bar{R}^0_{mn;\mu}\mathcal{M}^{\mu\lambda}_{ml} \sim O(1) $ for $ (l,\lambda) $ corresponding to the soft mode. Hence, the contribution of $ H_\mathrm{res} $ to the system dynamics is negligible when $\Omega\ll\Omega^*= \ell_{\mathrm{soft}}|G_\mathrm{NN}|$. For instance, using the values employed in Fig. 3 of the main text, one gets $ \Omega^*/h\sim 1 $ MHz. 

We now briefly comment on the infinite ion chain case. Here, by substituting the expanded potential of Eq.~\eqref{SM:eq:Vapprox} in $ H_\mathrm{int} $, we get
\begin{equation}\label{SM:eq:Hint}
	H_\mathrm{int}= \primesum_{m,n}V_R(\bm{r}_m^0,\bm{r}_n^0) n_m n_n+\sum_{\mu} \primesum_{m,n} \int_{-\pi}^{\pi} \mathrm{d}k\ W^{\lambda}_{mn}(k) \left[a^\dagger_{\lambda}(k) + a_{\lambda}(-k)\right] n_m n_n + O(\ell^2_\mathrm{min}/a^2),
\end{equation}
with $ W^{\lambda}_{mn}(k)=(2/\pi)^{1/2} \ell_{\lambda}(k) G_{mn}\bar{R}^0_{mn;\lambda}e^{i k m}$, where $G_{mn}$ and $\bar{R}^0_{mn;\lambda}$ are defined as in Eq.~\eqref{SM:eq:Hintfinite}. As in the finite chain case, the spin-phonon coupling term in the right hand side can be eliminated via the polaron transformation $ U=e^{-S} $~\cite{Deng:PRA:2005,Porras:2004,Gambetta:2020}, with 
\begin{equation}\label{SM:eq:polaron}
	S=-\sum_{\lambda}\int_{-\pi}^{\pi}\mathrm{d}k \left[\beta_\lambda(k) a^\dagger_{\lambda}(k)-\beta^*_\lambda(k)a_{\lambda}(k)\right]\quad\text{and}\quad \beta_\lambda(k)=\primesum_{m,n} \frac{W^\lambda_{mn}(k)}{\hbar \Omega_{\lambda}(k)} n_m n_n.
\end{equation}
This leads to a transformed Hamiltonian $ H=UHU^{-1}$ with the same form as Eq.~\eqref{SM:eq:H_CS}, in which the effective multi-body interaction coefficients contained in $ H_\mathrm{spin} $ are given by
\begin{equation}\label{SM:eq:Vmnij}
	\tilde{V}_{mnij}=\sum_{\lambda}\int_{-\pi}^{\pi} \mathrm{d}k\ \frac{W^\lambda_{mn}(k)[W^\lambda_{ij}(k)]^*}{\hbar \Omega_{\lambda}(k)}.
\end{equation}

\section{Microwave dressed potential in a trapped Rydberg ions chain}

Here, we provide further details about the derivation of the tailored microwave (MW) dressed interaction potential, shown in Fig. 2(b) of the main text, which allows to maximize the strength of the effective interactions.

\begin{figure}
	\centering
	\includegraphics[width=\textwidth]{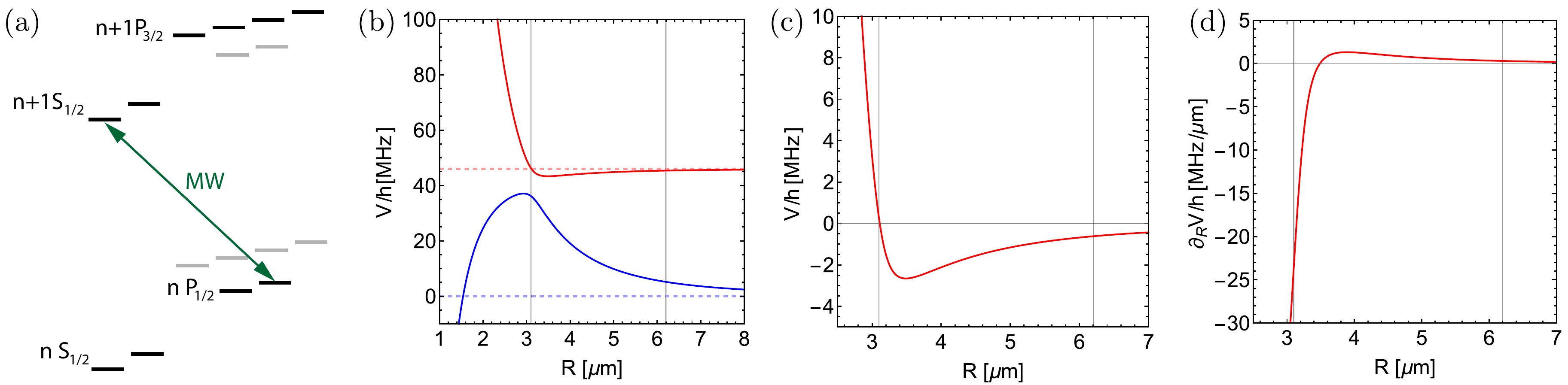}
	\caption{{\bf MW dressing scheme.}  (a) The four Rydberg states $ mS_{1/2} $, $ mP_{3/2} $, $ nS_{1/2} $, and $ nP_{1/2} $ (with $ m=n+1 $) are connected via dipole transitions. Two Zeeman sublevels of states $ nP_{1/2} $ and $ mS_{1/2} $ are coupled by a MW field with Rabi frequency $ \Omega_{\mathrm{MW}} $ and detuning $ \Delta_{\mathrm{MW}}. $ (b) Eigenvalues of $ H_\mathrm{MW} $ as a function of the ion-ion distance $ R $. Deviations of the latter from ion bare energies, highlighted by the corresponding dashed lines, measure the strength of the effective ion-ion interaction potential. Gray lines denote the distance between NNs, $ d_{\mathrm{NN}}=3.1\ \mu\mathrm{m} $, and NNNs, $ d_{\mathrm{NNN}}=2d_{\mathrm{NN}} $, corresponding to a linear chain configuration with a longitudinal trapping frequency $ \omega_x=2\pi\times 1.3 $ MHz (see main text for details). Here, $ \Delta E/h=46 $ MHz, $ \mathcal{C}_{3\mathrm{d}}/h= 1234\ \mathrm{MHz}\,\mu\mathrm{m}^3 $, and $ \mathcal{C}_{3\mathrm{f}}/h= -\mathcal{C}_{3\mathrm{e}}/h=150\ \mathrm{MHz}\,\mu\mathrm{m}^3 $.  (c) Effective potential obtained from the upper branch of panel (b) (red curve) as a function of the ion-ion distance $ R $. The values of the potential at NNs and NNNs are  $ V_\mathrm{NN}/h=0.25 $ MHz and $ V_\mathrm{NNN}/h=-0.6 $ MHz, respectively. (d) Gradient of the potential of panel (b) as a function of the ion-ion distance $ R $, with $ G_\mathrm{NN}/h=23.6\ \mathrm{MHz}/\mu\mathrm{m} $ and $ G_\mathrm{NNN}/h=0.3\ \mathrm{MHz}/\mu\mathrm{m} $. \label{SM:fig:potential}}
\end{figure} 

We consider a setup of $ {}^{88}\mathrm{Sr}^+ $ trapped Rydberg ions and focus on the four Rydberg states $ mS_{1/2} $, $ mP_{3/2} $, $ nS_{1/2} $, and $ nP_{1/2} $ (with $ m=n+1 $) shown in Fig.~\ref{SM:fig:potential}(a). For a pair of Rydberg excited ions, all these levels are coupled by the dipole-dipole interaction $V_\mathrm{dd} = \frac{1}{4\pi\epsilon_0} \Big[ \frac{\bm{\mu}_1\cdot\bm{\mu}_2 - 3(\bm{\mu}_1\cdot\bm{n})(\bm{\mu}_2\cdot\bm{n})}{|\mathbf{r}| ^3} \Big]$, with $ \bm{\mu}_i $ the electric dipole operator of ion $ i $, $ \bm{r} $ the distance between the two ions, and $ \bm{n}=\bm{r}/|\bm{r}| $. Moreover, two of the Rydberg states, namely $ mS_{1/2} $ and $ nP_{1/2} $, are coupled by a MW field. For a single two-level ion, the MW coupling Hamiltonian is
\begin{equation}
	H_\mathrm{MW}=\frac{1}{2}\begin{pmatrix}
		\Delta_\mathrm{MW} &\Omega_\mathrm{MW} \\
		\Omega_\mathrm{MW}  & -\Delta_\mathrm{MW}
	\end{pmatrix}.
\end{equation}
We denote the lowest energy dressed eigenstate of $ H_\mathrm{MW} $ by $ \ket{-}=\big(\ket{mS_{1/2}}-\ket{nP_{1/2}}\big)/\sqrt{2} $, with corresponding eigenenergy $E_{\ket{-}}=-\Omega_\mathrm{MW}/2$ (we set $E_{|mS_{1/2}\rangle}=0$). For a two ion system, in the regime $\Omega_\mathrm{MW}\gg V_\mathrm{dd}$, $V_\mathrm{dd}$ causes a distance-dependent energy shift $\bra{--}V_\mathrm{dd}\ket{--} = \frac{\mathcal{C}_{3\mathrm{e}}}{R^3}$ on the two-ion state $\ket{--}$, whose eigenenergy becomes $E_{\ket{--}}=-\Omega_\mathrm{MW} + \frac{\mathcal{C}_{3\mathrm{e}}}{R^3}$. Note that the energy offset can be tuned via the MW Rabi frequency.

The other two states, $ mP_{3/2} $ and $ nS_{1/2}$, are also coupled by the dipole-dipole interaction $V_\mathrm{dd}$. The corresponding Hamiltonian for a two-ion system in the $ mP_{3/2} $, $ nS_{1/2}$ basis is
\begin{equation}\label{SM:eq:mPnS}
	H_\mathrm{d}=\begin{pmatrix}
		-2E_{\mathrm{SP}} & 0 & 0 & 0 \\
		0 & 0 & \frac{\mathcal{C}_{3\mathrm{d}}}{R^3} & 0 \\
		0 & \frac{\mathcal{C}_{3\mathrm{d}}}{R^3} & 0 & 0 \\
		0 & 0 & 0 & 2E_{\mathrm{SP}}
	\end{pmatrix},
\end{equation}
where $E_{\mathrm{SP}}$ is the energy separation between $ nS_{1/2}$ and $ mP_{3/2} $ and the average of $ nS_{1/2}$ and $ mP_{3/2} $ energies is set to 0. Here, $\mathcal{C}_{3\mathrm{d}}/R^3 = \bra{nS_{1/2}mP_{3/2}}V_\mathrm{dd}\ket{mP_{3/2}nS_{1/2}}$ is the coupling caused by dipole-dipole interaction. The two-ion eigenstates of Eq.~\eqref{SM:eq:mPnS} are $\ket{nS_{1/2}nS_{1/2}}$, $\ket{SP_+}=\big(\ket{nS_{1/2}mP_{3/2}} + \ket{mP_{3/2}nS_{1/2}}\big)/\sqrt{2}$, $\big(\ket{nS_{1/2}mP_{3/2}} - \ket{mP_{3/2}nS_{1/2}}\big)/\sqrt{2}$, and $\ket{mP_{3/2}mP_{3/2}}$. Note that their eigenenergies depend on distance because of the dipole-dipole interaction ($\sim\mathcal{C}_{3\mathrm{d}}/R^3$).

By acting on the MW Rabi frequency, the energy of the $\ket{--}$ state can be tuned close to that of $\ket{SP_+}$. If we use the energy of $\ket{SP_+}$ for $R\rightarrow\infty$ as the zero energy reference, the energy of $\ket{--}$ for $R\rightarrow\infty$ becomes $\Delta E$, where the energy difference $ \Delta E $ between the states $\ket{--}$ and $\ket{SP_+}$ can be tuned via the MW field. Furthermore, these two two-ion states are coupled by the dipole-dipole interaction $V_\mathrm{dd}$, with $\bra{--} V_\mathrm{dd} \ket{SP_+} = \mathcal{C}_{3\mathrm{f}}/R^3$. Hence, the effective overall Hamiltonian for a two-ion system in the $\ket{--}$, $\ket{SP_+}$ basis is
\begin{equation}\label{SM:eq:HMW}
	H_\mathrm{MW}=\begin{pmatrix}
		\frac{\mathcal{C}_{3\mathrm{d}}}{R^3} &\frac{\mathcal{C}_{3\mathrm{f}}}{R^3} \\
		\frac{\mathcal{C}_{3\mathrm{f}}}{R^3}  & \Delta E+\frac{\mathcal{C}_{3\mathrm{e}}}{R^3} 
	\end{pmatrix}.
\end{equation}
The corresponding eigenenergies as a function of the ion-ion distance $ R $ are plotted in Fig~\ref{SM:fig:potential}(b). The combination of dipole-dipole interactions and MW-tuned F\"{o}rster resonance allows to conveniently tailor the effective ion-ion interaction energy, given by the difference between the eigenenergies of Eq.~\eqref{SM:eq:HMW} and the bare energies $ 0 $ and $ \Delta E $ of the states $\ket{SP_+}$ and $\ket{--}$, respectively. In particular, as discussed in the main text, effects of the multi-body interactions arising from the interplay between electronic and vibrational degrees of freedom are maximized in the presence of a Rydberg interaction potential with weak NNs and NNNs interactions ($ V_\mathrm{NN} $ and $ V_\mathrm{NNN}  $, respectively) but large values of the corresponding gradients (i.e., $ G_\mathrm{NN}  $ and $ G_\mathrm{NNN} $). As shown in Fig~\ref{SM:fig:potential}(c) and (d), such a potential can be obtained by the dressing scheme discussed so far in state-of-the-art setups of trapped Rydberg ions. 

\end{document}